\documentclass[twocolumn]{aastex631}

\usepackage{natbib}

\usepackage{amsthm}
\usepackage{graphicx}
\usepackage[space]{grffile}
\usepackage{latexsym}
\usepackage{amsfonts,amsmath,amssymb}
\usepackage{url}
\usepackage[utf8]{inputenc}
\usepackage{fancyref}
\usepackage{hyperref}
\hypersetup{colorlinks=true,citecolor=blue,urlcolor=blue,pdfborder={0 0 0},}
\usepackage{textcomp}
\usepackage{multirow,booktabs}
\usepackage{caption}
\usepackage{subcaption}
\usepackage{tablefootnote}
\usepackage{xspace}

\shorttitle{Density Variations}
\shortauthors{Berardo \& de Wit 2022}

\begin{document}
\title{Tidal Distortions as a Bottleneck on Constraining Exoplanet Compositions}

\author{David Berardo}
\altaffiliation{Department of Physics and Kavli Institute for Astrophysics and Space Research, Massachusetts Institute of Technology, Cambridge, MA 02139, USA}
\altaffiliation{FRQNT Doctoral Research Scholarship}

\author{Julien de Wit}
\altaffiliation{Department of Earth, Atmospheric and Planetary Sciences, Massachusetts Institute of Technology, Cambridge, MA 02139, USA}

\correspondingauthor{David Berardo}
\email{berardo@mit.edu}

\begin{abstract}
    Improvements in the number of confirmed planets and the precision of observations implies a need to better understand subtle effects which may bias interpretations of exoplanet observations. One such effect is the distortion of a short period planet by its host star, affecting its derived density. We extend the work of \cite{burton:2014,correia:2014} and others, using a gravitational potential formulation to a sample of nearly 200 planets with periods less than three days. We find five planets exhibiting density variations of $>$ 10\%, and as many as twenty planets with deviations $>$ 5\%. We derive an analytic approximation for this deviation as a function of the orbital period, transit depth, and mass ratio between the planet and host star, allowing for rapid determination of such tidal effects. We find that current density error-bars are typically larger than tidal deviations, but that reducing the uncertainty on transit depth and RV amplitude by a factor of three causes tidal effects to dominate density errors ($>50\%$) in $>$40\% of planets in our sample, implying that in the near future upgraded observational precision will cause shape deviations to become a bottleneck with regards to analysis of exoplanet compositions. These two parameters are found to dominate uncertainties compared to errors on stellar mass and radius. We identify a group of eight planets (including WASP-19 b, HAT-P-7 b, and WASP-12 b) for which current density uncertainties are as much as four times smaller than the potential shift due to tides, implying a possible 4$\sigma$ bias on their density estimates.
\end{abstract}


\section{Introduction}

As the list of confirmed exoplanets grows we continuously expand the sampled space of known planetary parameters. Categories of planets such as those with ultra-short orbital periods have gone from containing a handful of planets to hundreds of planets thanks to missions such as \textit{Kepler} \citep{borucki:2010} and \textit{TESS} \citep{ricker:2014}. In addition to this increase in population, the precision of instruments has continued to reach new heights, reducing the uncertainty in quantities such as transit depth or planetary mass. This trend will accelerate further with the next generation of observatories and instruments such as JWST and PLATO \citep{heras:2020}, as well as high precision RV instruments such as CARMENES \citep{reiners:2018} and ESPRESSO \citep{schmidt:2021}.  This increase in both the size and quality of our sample implies that subtle effects which in the past where either too small to be detectable or which affected a single digit number of planets may no longer be disregarded. An example of this behaviour is the `Transit Light Source' effect \citep{rackham:2018}, in which variability of the stellar surface causes biases in atmospheric characterisation by mimicking or muting effects which produce similar results, acting a bottleneck towards properly understanding a planets atmosphere. 

The focus of this work is on effects which alter the shape of an exoplanet, which is often considered to be a perfect sphere such as in the commonly used models of \cite{mandel:2002}, implemented in the widely used batman package \citep{kreidberg:2015b}.
\begin{figure*}[!ht]
	\centering
	\includegraphics[scale=0.4]{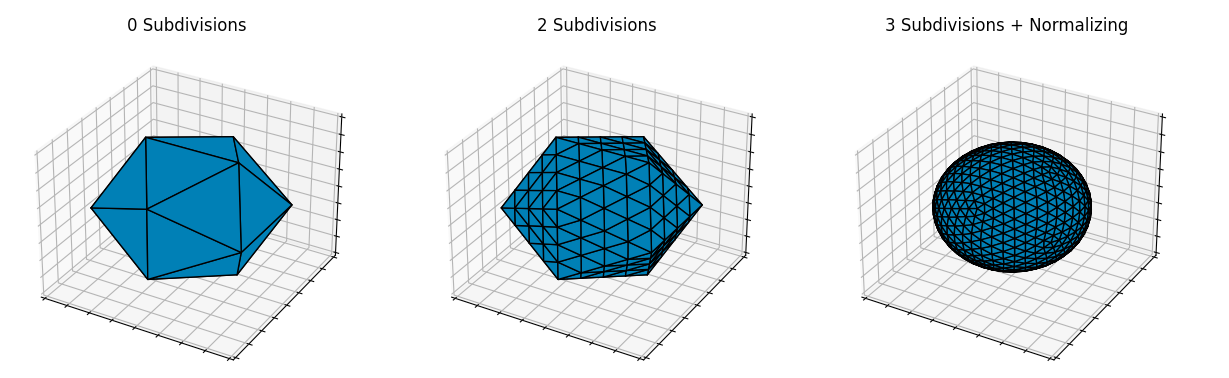}
	\caption{An illustration of the process by which the surface of the sphere is constructed. Starting from an icosahedron on the left, triangular faces are continually subdivided. Finally, the points are normalized to generate a uniformly sampled sphere.}
	\label{fig:icosphere}
\end{figure*}
For short period planets close to their host star, one such effect are tidal distortions which can cause a planet to bulge out towards its host star \citep{leconte:2011}. This effect in particular has the potential to introduce a significant bias on the density of a planet since its sky projection remains close to a perfect circle. When considering for example a planet which is deformed due to rotation causing ts equator to bulge, its projection becomes elliptical \citep{seager:2002,barnes:2003}. In this case, subtle difference in the shape of ingress / egress of the transit lightcurve may be used to break the degeneracy between a spherical and oblate planet \citep{carter:2010a,berardo:2022a}. For tidally deformed planets, phase curve observations which observe the planet from different directions could in principle determine these so called `ellipsoidal variations' through lightcurve deviations \citep{correia:2014,kreidberg:2018}, however full phase curve observations require a significant amount of observing time to obtain, and at high precision there is likely to be a significant amount of degeneracy between the orbit, shape, and brightness distribution of a planet \citep{dewit:2012}.

Tidal distortions imply an underestimate of the volume of a planet, which in turn implies an overestimate of its bulk density. Theoretical considerations of the effect of this have previously been studied in \cite{leconte:2011} This effect has already been considered, primarily in the work of \cite{burton:2014}, which calculated the magnitude of the distortion and the degree to which it altered the density measurement for a sample of just over 30 planets. Additionally, 
\cite{correia:2014} expanded on this work using a more detailed model to derive an analytic expression for the change in density as a function of distance to the host star.

In this work we aim to expand on these efforts in several ways. Our primary effort is to increase the sample of planets analysed using a gravitational potential model, which has been found to provide similar results to more complicated structural models. In the time since these previous studies were published, roughly 6x as many planets have now been found to be in the space of parameters which are susceptible to tidal distortion effects (i.e. planets with orbital periods below three days on circular orbits).

In section \ref{sec:model description} we briefly outline the theory of tidal deformation and describe our method for calculating the effects of tidal interactions, and thus altered planetary densities. In section \ref{sec:results} we first highlight our sample of planets to be analysed, followed by the results of our analyses. We highlight trends as a function of various system parameters and derive an approximation which accurately describes the changes in density without the need for a full simulation. In section \ref{sec:discussion} we first highlight the biases that may be introduced when attempting to retrieve the interior composition of a planet using mass-radius relations under the assumption of being perfectly spherical. We then compare the changes in density to current density uncertainties, and we also analyse the relative contributions to these uncertainties from five parameters underlying parameters. This allows us to determine how upcoming improvements in quantities such as planet mass and stellar parameters will affect the ability to ignore such effects, for example through extreme precision radial velocity efforts \citep{eprv:2021}.

\section{Calculating the Density of a Tidally Deformed Planet}
\label{sec:model description}
\subsection{Physical description of scenario}
\label{sec:theory}

To model the shape of the planet, we follow a similar methodology as that of \cite{burton:2014}, where the surface of the planet is assumed to be on a gravitational equipotential. The value of the gravitational potential generated by a rotating planet and its host star is calculated using the Roche approximation \citep{chandrasekhar:1987}:

\begin{align}
\label{eq:roche_pot}
    \Phi_1 &= -\frac{GM_1}{\left((x+a)^2+y^2+x^2\right)^{1/2}}\\
    \Phi_2 &= -\frac{GM_2}{\left(x^2+y^2+x^2\right)^{1/2}}\\
    \Phi_3 &= -\frac{1}{2}\Omega^2\left[(x+\mu_1 a)^2+y^2\right]
\end{align}
\begin{figure*}[!ht]
	\centering
	\includegraphics[scale=0.35]{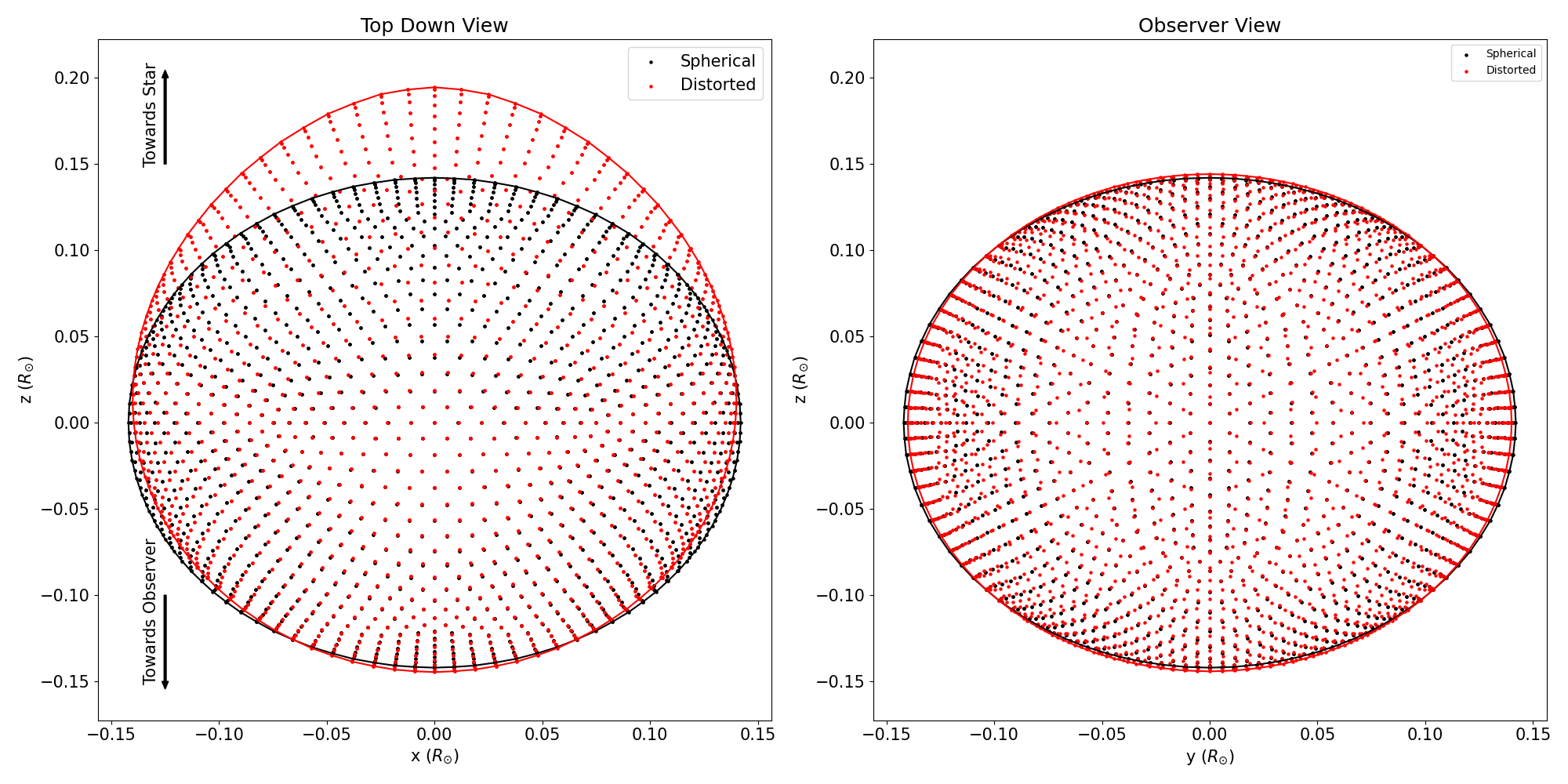}
	\caption{This figure shows two views of the surface of WASP-19 b. Points in black show the spherical planet which matches the observed transit depth, while points in red show the surface generated by fitting for an equipotential while also matching the observed transit depth. On the left we see a top down view of the orbital plane. On the right we see the view along the line of sight between the centers of mass of the planet and star.}
	\label{fig:surface variation}
\end{figure*}
where $G$ is the gravitational constant, $M_1$ is the mass of the host star, $M_2$ is the mass of the planet, $a$ is the separation between the host star and planet (i.e. the semi-major axis of a circular orbit), $\mu_1 = M_1 / (M_1+M_2)$ and $\Omega = 2\pi/P$ where $P$ is the orbital period of the planet. The coordinate system is such that the origin is placed at the center of the planet. The x coordinate points along the line connecting the center of masses of the two bodies, the z axis points along the orbital plane in the direction of motion of the planet, and the y axis points normal to the orbital plane. 


In order to use such an approximation to model the distortion of a planets surface, we assume the planet is both tidally locked as well on a non-eccentric orbit. As we shall see in later sections, the effect of the distortion is strongest for low period planets (p $<$ 3 days) which are most likely to be tidally locked and be on circular orbits\citep{barnes:2017}. 

\subsection{Calculating the volume of a deformed planet}
\label{sec:method}

We first calculate the surface of a deformed planet and then `measure' its volume in order to determine the amount by which its density is altered. In order to generate the surface of our planet, we first construct a geodesic icosahedron as an approximation of a sphere. This is an object commonly used in computer graphics and 3D rendering software which has the benefit of having its points uniformly spread out across its surface. We begin with the vertices of an icosahedron and then iteratively subdivide each of its faces into smaller triangles (as shown in figure \ref{fig:icosphere}). After the last round of subdivisions we normalize the length of each vertex from the origin to generate a tiled sphere.

This process leaves us with a collection of triangular faces which allows us to calculate two necessary quantities, the total projected surface area visible to an observer as well as the enclosed volume of each tetrahedron generated by the origin and any given triangular face. An additional benefit of this method is that we can adjust the number of iterations in order to achieve any level of precision we desire. We find that after 5 subdivisions the calculated volume of our icosphere differs from that of a perfect sphere by only 0.05\%, while the calculated projected area varies by only 0.03\%. We use this as a benchmark for the accuracy of our method and fix all further calculations to 5 subdivisions, which gives us a surface of 10242 triangular tiles.

We next scale each vertex radially until all points have the same gravitational potential, which requires us to pick a value of the equipotential $\Phi$. We choose $\Phi$ such that the projected surface area matches the observed transit depth, similar to what is done in \cite{burton:2014}. We first evaluate the equipotential function for a range of radii centered on the spherical planet radius. For each value of $\Phi$ generated this way, we then calculate the radius of each vertex using a least squares regression in order to find the surface of constant potential. For this surface, we then calculate the projected planet area. This gives us a mapping between gravitational potential and transit depth, which we use to select the value of $\Phi$ which corresponds to any depth value of our choosing.

The result of this process is shown in figure \ref{fig:surface variation}, where we have calculated the deformation of WASP-19 b \citep{hebb:2009} using the described process. This example highlights the potential for tidal deformation to alter a planets measured density. In the left panel we see a significant deviation from a pure sphere, as the planet is pulled towards its host star. However in the right panel we see that the observer-projected shape of the planet remains nearly perfectly circular.

\section{Density Variations of Confirmed Planets}
\label{sec:results}

\subsection{Planet Sample}
\label{sec:sample}
We begin with the full list of confirmed planets found in the exoplanet archive \citep{confirmedplanets} which currently contains just over 5000 exoplanets. As mentioned in the previous section, as well as motivated by the results of \cite{burton:2014}, we focus our efforts on short period planets, specifically planets with orbital period of less than 3 days. We do also analyze planets with periods in the range of 3-5 days, but those were found to have negligible tidal distortion effects, consistent with expectations. 

We additionally focus only on planets which have reported mass values. In principle, relative variations in density can be measured based on just changes in planet volume which is the focus of this work. However we also consider the magnitude of such a difference relative to the uncertainty in the measured density, for which a mass value (along with an error-bar) is required. We also cut for planets with eccentricity values below e = 0.05. This leaves us with a final sample of 196 planets, just over 6 times larger than the sample of planets used in \cite{burton:2014}.


\subsection{Density Variation Results}
\label{sec:density variations}

We apply the process described in section \ref{sec:method} to each of the planets in our sample. For each planet, we calculate its volume under tidal deformation that produces a depth value which matches the median reported value to within 0.1\% in order to minimize differences caused by truncation or any other numerical effects. All analyses in this section use these values in order to compare the spherical and tidal planet densities. 
\begin{figure}[!ht]
	\centering
	\includegraphics[scale=0.38]{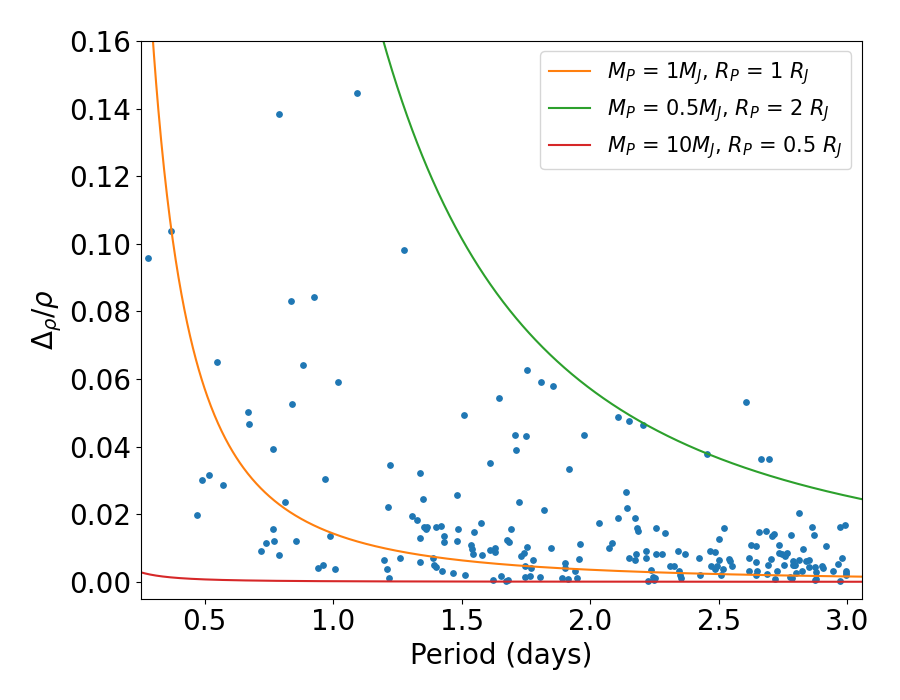}
	\caption{Relative change in the density of planets with orbital periods less than three days. The curves show the functional dependence of equation \ref{eq:functional form} for representative values of planet mass and radius.}
	\label{fig:period dependance}
\end{figure}

\subsubsection{Absolute changes in density \& trends}
\label{sec:measured density changes}
We first look at the percent difference in the density of each planet under the assumptions of being perfectly spherical or tidally deformed

\begin{equation}
\frac{\Delta\rho}{\rho_{sph}} = \frac{\rho_{sph} - \rho_{tide}}{\rho_{sph}}
\end{equation}

where we calculate the value of $\rho_{sph}$ ourselves using the reported values of mass, depth, and stellar radius. This is done to ensure a fair comparison in order to accurately represent the amount by which density can shift due to changes in volume. As we will see in the next section, this quantity is often comparable to the density uncertainty, which is set by the underlying uncertainties from transit depth and RV semi-amplitude measurements used to calculate density.  Using the reported value of planet density thus suffers the risk of including measurement uncertainty (depending on how density is reported which varies between analyses) when at this stage we only wish to determine intrinsic differences. Thus we ensure that both our measurements correspond to identical values of depth and planet mass.

The results of this are shown in figure \ref{fig:period dependance}. We show the variation as a function of orbital period, where we note that the variation decreases as period increases. This is not surprising given the factor of $1/p^2$ which appears in the potential equation, and acts as an additional confirmation that our code is accurately calculating planet deformations (a similar trend with fewer planets was also seen in \cite{burton:2014}).

We truncate the plot at an upper limit of $p = 3$ days, but note that we calculated the deformation out to a period of 5 days and found that the trend continued, in particular the upper envelope which flattens out at a maximal deviation of $\sim$ 2$\%$. We find that for planets with orbital periods below 1.5 days, the tidal density may deviate by as much as 15\% compared to the density which comes from assuming a perfectly spherical planet.

\subsubsection{Functional Approximation of Density Variations}
\label{sec:functional form}
The scatter in figure \ref{fig:period dependance} implies that orbital period is not the sole factor in determining density variations, which is also apparent from the additional terms in equation \ref{eq:roche_pot}. We attempt to derive a functional form of the variation in density by comparing the full tidal potential to that of an isolated spherically symmetric body, given by $\Phi_{sph} = -GM_p/r$. We first assume that points along the surface of a tidally distorted planet are at a similar distance to the center of the planet as for a non-distorted planet, i.e. no part of the planet is distorted by a factor of say two or more. Thus in the Roche approximation we may replace quantities such as $x^2+y^2+z^2$ with $r_p^2$ or equivalently $\sqrt{\delta}R_s$ where $\delta$ is the observed transit depth and $R_s$ is the stellar radius. We also assume that $a >> R_p$ (in our sample we always have at least $a/R_p > 10$), and also that that solar mass is much larger than the planetary mass (for our sample we always have $M_s/M_p > 10^2$). Under these assumptions the three terms from equation \ref{eq:roche_pot} become: 

\begin{equation}
    \Phi_1 \sim -\frac{GM_s}{a},\
    \Phi_2 \sim -\frac{GM_p}{r_p},\ 
    \Phi_3 \sim -\frac{1}{2}\Omega^2a^2
\end{equation}

which we then combine and scale by $\Phi_{sph}$ to get
\begin{equation}
    \frac{\Phi_{sph} - \Phi_{tide}}{\Phi_{sph}} = \frac{3}{2}\frac{M_p}{M_s}\frac{r_p}{a}
\end{equation}

where we've used Kepler's 3rd law to combine the orbital period and semi-major axis terms.

We now have an equation for the change in gravitational potential, which must be converted to a change in density. Given that the potential is treated as a radial 1D function, a reasonable assumption might be that the scaling term (rp/a) needs to be cubed in order to obtain a relationship for density. To confirm this, we parameterize the change in density as 

\begin{equation}
    \frac{\Delta\rho}{\rho_{sph}} = \alpha \left(\frac{Mp}{Ms} \right)^\beta\left(\frac{r_p}{a} \right)^\gamma
\end{equation}

and fit for $\alpha$, $\beta$, $\gamma$ against the calculated values for $\Delta_\rho / \rho$. We do indeed find that $\gamma \sim 3$, as well as $\alpha\sim2$ and $\beta\sim1$. We present the final effect on the change in density (having re-converted to orbital period) as 

\begin{equation}
    \label{eq:functional form}
    \frac{\Delta\rho}{\rho_{sph}} = 0.01428\left(\frac{P}{\mathrm{day}}\right)^{-2}
    \left(\frac{R_p}{\mathrm{R_J}}\right)^{3}
    \left(\frac{M_p}{\mathrm{M_J}}\right)^{-1}
\end{equation}

We plot this function for representative values of planet radius and planet mass in our sample in figure \ref{fig:period dependance}, where we find good agreement particularly in the upper envelope of the data points which closely follows an inverse square dependence on the period.

\begin{figure}[!ht]
	\centering
	\includegraphics[scale=0.4]{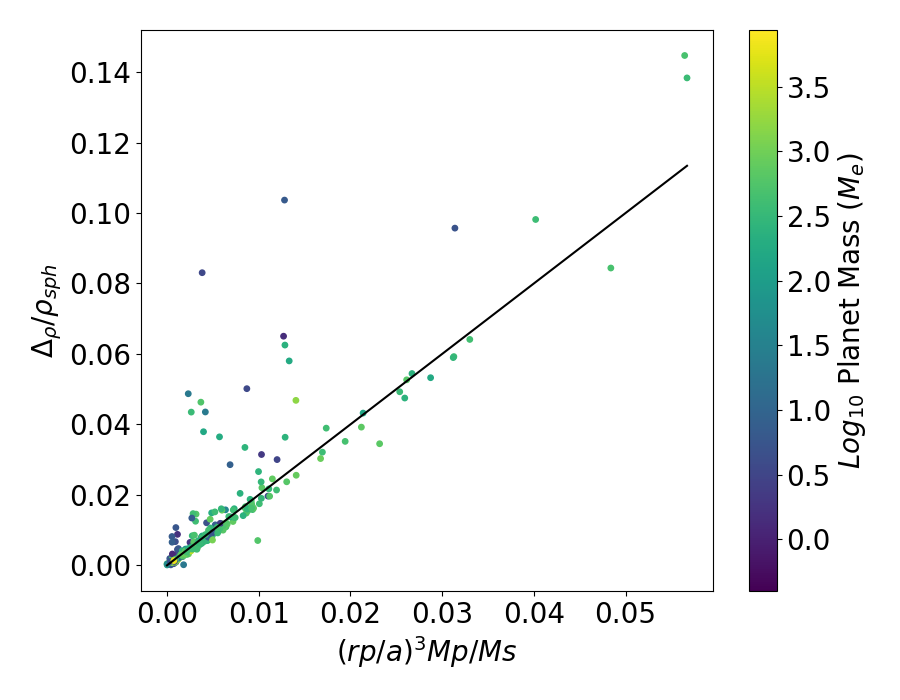}
	\caption{We show here agreement between the functional form the of the density perturbation we derive in section \ref{sec:functional form} (x-axis) against the values calculated in section \ref{sec:measured density changes} (y-axis). The black line a linear relationship with a slope of 2 passing through the origin. The coloring represents the mass of each planet on a log scale.}
	\label{fig:functional agreement}
\end{figure}
\begin{figure*}[!ht]
	\centering
	\includegraphics[scale=0.48]{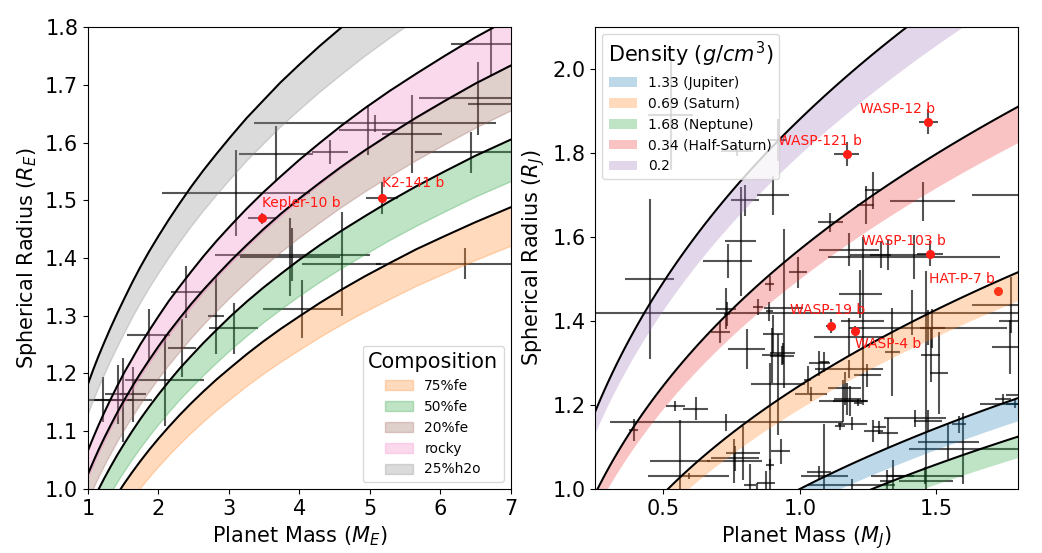}
	\caption{Mass-Radius relationships along with data points (and error-bars) for the sample of planets considered in this work. The left and right panels refer to low and high mass planets respectively. Black curves in the left figure are taken from \cite{zeng:2019}. Colored bands represent variations of these curves by up to $4.7\%$ in radius, corresponding to density variations of up to $15\%$. The right hand panel shows a similar phenomenon for high mass planets, where we show constant-density relations corresponding to solar system gas giants giant densities, as well as a planet with half the density of Saturn and a planet with a density of $0.2g/cm^3$ representative of `super puffs'. The planets highlighted in red are those mentioned in Table \ref{tab:planets with small sigmas} whose density uncertainty is less than the deviations caused by tides.} 
	\label{fig:mr curves}
\end{figure*}
We additionally compare this analytic description of the change in density directly to the values calculated in section \ref{sec:measured density changes} and show the results in figure \ref{fig:functional agreement}. We find that the bulk of the data points follow a linear relationship with a slope of $\sim$2, although we do still note a certain amount of scatter above the line. Planets which deviate significantly from the trend tend to have smaller masses (closer to being earth sized). This implies that one or more of the assumption we have made in deriving this relationship breaks down for sufficiently low mass planets. At the scales involved, our approximation deviates by at most a factor of ten from the true relative density change. This implies an underestimate of the true volume change by at most 10\%, which in turn corresponds to an error on the linear scale of the planet by $~1.1^{1/3}\sim3\%.$ The true deviation is almost always larger than our functional approximation. Thus equation \ref{eq:functional form} represents a fairly robust metric to determine if a planet may be susceptible to tidal deformations, without needing to run a full gravitational potential calculation.

A similar metric was derived in \cite{correia:2014} (eq. 27), using a different approach considering the Love number and fluid displacement of an exoplanet \citep{love:1911}. The result they obtain is similar in that it is proportional to the ratio of planet to stellar mass, as well to the third power of planet size to orbital semi-major axis. While we find a constant scaling factor of two, they obtain a scaling factor of $7h_f/4$, where $h_f$ is the fluid second Love number. Estimating $h_f$ using the Darwin-Radau relation \citep{bourda:2004} and a value of $\sim0.27$ for the moment of inertia of Jupiter \citep{ni:2018} gives a prefactor of $2.5$. This difference of $~25\%$ in estimated tidal density is well below the measurement uncertainty on planet density, and using either equation would indicate weather or not the planet of density may be significantly different from that of a spherical planet.

\begin{figure*}[ht!]
	\centering
	\includegraphics[scale=0.5]{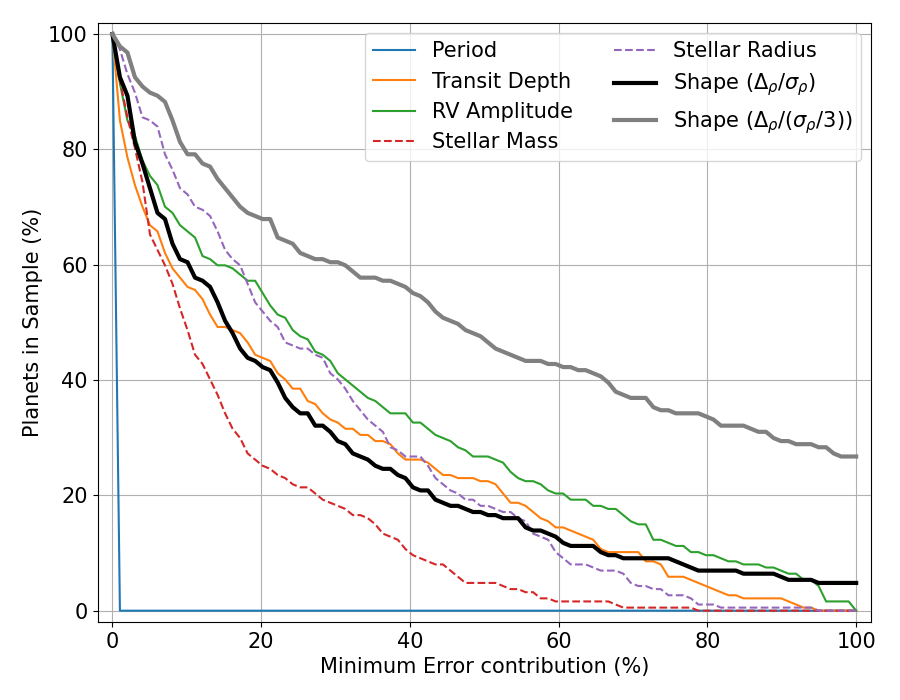}
	\caption{This figure illustrates the relative contribution of five underlying factors to the derived measurement error of a planets density. The x-axis represents a minimum amount that a given parameter contributes to the overall uncertainty on density. Solid colored lines represent directly observable quantities (period, transit depth and RV amplitude), while dashed colored lines refer to model-dependant quantities (stellar mass and radius).  The black line (`shape') shows the ratio of tidally-induced variation to measurement error ($\Delta_\rho / \sigma_\rho$). The grey line shows a similar value, after having artificially reduced the overall uncertainty on density by a factor of three to highlight the effect of future measurement improvements.}
	\label{fig:error breakdown}
\end{figure*}

An additional consideration of \cite{correia:2014} is the effect of inclination on the derived density, which introduces a correction term in their equation 27 proportional to $cos^2i$. We find that for the planets in our sample, the effect of this correction term is at most $2\%$ for a handful of planets, and more typically well below a $1\%$ correction. Thus in our analysis we have chosen to neglect the effects of inclination, in order to provide a simple framework which still captures the bulk of the deviations. Even when considering the maximum inclination that would allow for a transit to occur, the correction term is at most $2\%$ for the shortest period planets in our sample, and for most planets is much less than $1\%$.

\section{Discussion}
\label{sec:discussion}
In the previous section we considered the absolute changes in density a planet may experience under the effects of tidal forces. We now focus on contextualizing these results with regards to measurement accuracy, and biases that may be introduced in measuring planet compositions to high precision.

\subsection{Uncertainty in Mass-Radius Relations}
During transit a tidally distorted planet will still have a nearly circular projection while having a larger than expected volume as shown in figure \ref{fig:surface variation}. The implication of this is that a spherical transiting planet could have the same density as a deformed planet with a smaller projected area, due to the `hidden' extra volume. Thus when considering mass-radius composition curves, there is in-fact a degeneracy wherein a single curve could actually correspond to a range of projected radii, which we show in figure \ref{fig:mr curves}. The implication of this is that even if a planet had no error whatsoever on its transit depth, there would still remain uncertainty on its composition due to a lack of knowledge of its shape, becoming a bottleneck when attempting to measure planetary compositions to high precision.

We separate our sample of planets into low mass (Earth-sized) and high mass (Jupiter-size) planets, and for each we show a selection of various composition curves. For the low mass planets we show curves taken from \cite{zeng:2019}, for a range of iron fractions as well as an Earth-like composition and a planet with a 25\% water composition. For gas giant planets, we show a range of densities corresponding to the solar system gas giants, as well as a lower density of $0.2g/cm^3$ as a representative value of large planets with low densities, so called `super-puffs'\citep{masuda:2014,lopez:2014}. For each curve, we a plot a range of values (the colored regions) corresponding to a radius difference of $\sim5\%$, which corresponds to a maximal density variation of $\sim15\%$. This represents the range of projected radii which could all correspond to the same density. 

The effect of these considerations is that, for example, a composition of 20\% iron and one of pure-rock become a near continuous region of parameter space, and a planet such as Kepler-10 b, which we note in the next section has a relatively low measurement error, could now be equally described by either model. We additionally see planets which fall between models of 25\% water and one of pure rock. While their own uncertainties make the distinction clear, with one model being two or three standard deviations away, it becomes much less obvious which model is correct once the additional uncertainty from shape variations (colored bands) is considered.

\subsection{Uncertainty of Density Measurements}
In the previous section we considered the limiting case of perfect transit depth and mass radius knowledge and their effect on compositional analysis. We now focus on current measurement errors, how they compare to changes induced by tidal variations, and how upcoming improvement in precision of the quantities used to calculate density, namely transit depth, stellar radius, and planet mass (which itself depends on the stellar mass, RV semi-major amplitude, and orbital period) will in turn affect the uncertainty on density.

For a function $f$ which depends on independent variables $x_{i}$, we can write the uncertainty of $f$ (denoted $\sigma_f$) as:

\begin{equation}
    \sigma_f^2 = \sum_{i} \left(\frac{\partial f}{\partial x} \sigma_{x_i}  \right)^2  
\end{equation}

which for a density calculated using planet mass ($M_p$), transit depth ($\delta$) and stellar radius ($R_s$) becomes

\begin{equation}
    \label{eq:rho error}
    \sigma_{\rho} =  \sqrt{\left(\frac{\rho}{M_p}\sigma_{M_p}\right)^2
    +     \left(\frac{3}{2}\frac{\rho}{\delta}\sigma_{\delta}\right)^2
    + \left(3\frac{\rho}{R_s}\sigma_{R_s}\right)^2}
\end{equation}

 We note that most of the planets in our sample have reported values for their density along with an error-bar in their entries in the exoplanet archive. We additionally calculate the uncertainty by ourselves using equation \ref{eq:rho error} and the reported uncertainties for the involved quantities, and find a good agreement between the two values. An additional consideration is that the planet mass itself is dependant on the radial velocity semi-major amplitude (K), stellar mass ($M_s$), and orbital period (p), which allows us to write the uncertainty on the planet mass as:
 
 \begin{equation}
    \label{eq:mp error}
    \sigma_{M_p} =  \sqrt{\left(\frac{M_p}{K}\sigma_{K}\right)^2
    +     \left(\frac{3}{2}\frac{M_p}{M_s}\sigma_{M_s}\right)^2
    + \left(\frac{1}{3}\frac{M_p}{P}\sigma_{P}\right)^2}
\end{equation}

The benefit of calculating the uncertainty directly in this way is that we are then able to compare the relative contribution of each term to the overall uncertainty. We quantify the relative contribution of a variable $x_i$ as:

\begin{equation}
    \left(\frac{\partial \rho}{\partial x_i}\sigma_{x_i}\right)^2 / \sigma_\rho ^ 2
\end{equation}

such that the sum of the contributions of each variable is 100$\%$. The results of this breakdown are shown in figure \ref{fig:error breakdown}, where we illustrate how often a given parameter contributes a minimum amount to the uncertainty. We find for example that in our sample of planets the orbital period never contributes more than $0.0001\%$ relative to the other parameters, which is unsurprising given that the orbital period of a transiting planet is typically measured to extremely high precision.

\begin{table}[]
\begin{tabular}{c|ccc}
\hline
Error Contributor&Min \%&Max \%&Median \%\\
\hline
1&31&100&59\\
2&0&45&25\\
3&0&28&9\\
4&0&18&3\\
\hline
\end{tabular}
\caption{Summary of the ranked contributions to density error across all planets where 1 = largest contributing factor and 4 = smallest. We find that the largest source of error (regardless of which underlying parameter it comes from) comprises anywhere from 31\%-100\% of the uncertainty on a planets density, with a median value of 59\%.}
\label{tab:error breakdown}
\end{table}

For the remaining four parameters we can categorise them as being either measurement dependant (transit depth and RV amplitude) or model dependant (stellar mass and stellar radius). We find that it is the measurement parameters which more often contribute the largest amount of uncertainty, with the RV amplitude alone contributing at least 60\% of the relative uncertainty for $\sim$20\% of the planets in our sample, and in some case it even contributes almost the entirety of the uncertainty. Transit depth similarly can contribute as much as $90\%$ in some cases, whereas the model dependant parameters never contribute more than 80\% of the relative uncertainty. In table \ref{tab:error breakdown} we show the ranked breakdown of error contributions, in order of largest to smallest contributor (regardless of which parameter it comes from). We find for example that for a given planet the largest source of uncertainty always contributes at least 31\% of the error and potentially the entire uncertainty, with a median contribution of 59\%. This indicates that for most planets there is a single parameter which contributes more than half of the density uncertainty.

When we consider the overall uncertainty on the spherical density, we find that for most planets the calculated difference between the spherical and tidal density ($\Delta_\rho$) is smaller than the uncertainty ($\sigma_\rho$), illustrated by the black line in figure \ref{fig:error breakdown} which shows the ratio between the two.  This implies that with current data precision assuming a planet to be spherical in most cases does not introduce a significant statistical bias, but may be causing density error uncertainties to be underestimated. 

\begin{figure}[!ht]
	\centering
	\includegraphics[scale=0.4]{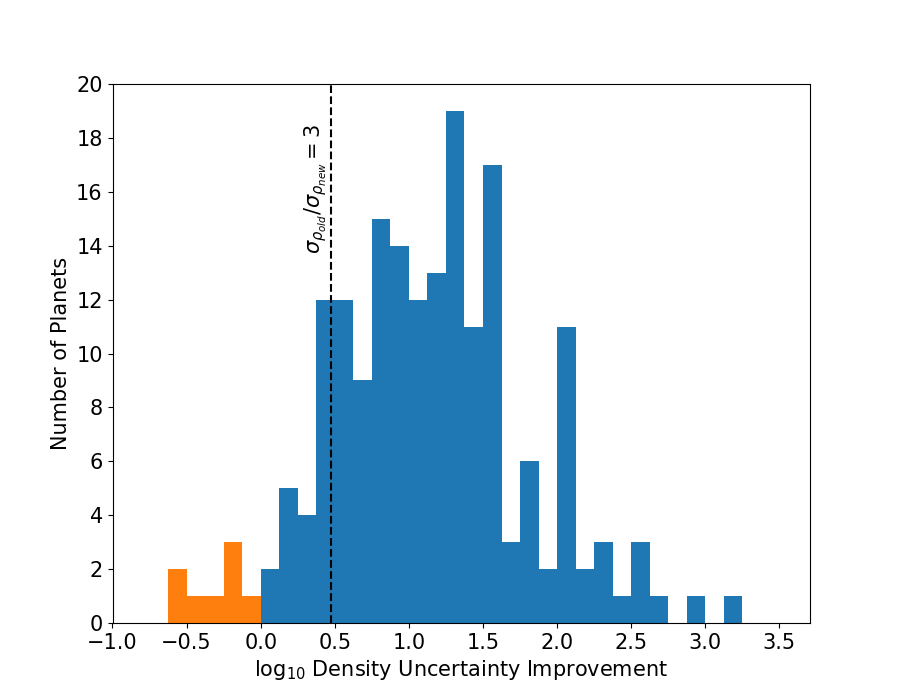}
	\caption{The factor by which the uncertainty on a planets density needs to be improved such that it is equal to the change in density due to tidal deformations. The black dotted line highlights a decrease by a factor of 3 in the uncertainty on spherical planet density. The orange region highlights planets whose density uncertainty is currently less than the difference between the spherical and tidal density values.}
	\label{fig:density uncertainty improvement}
\end{figure}

Given this result, we can then ask by how much the error on planet density needs to be reduced before we have $\sigma_{\rho} = \rho_{sphere} - \rho_{tidal}$, which we show in figure \ref{fig:density uncertainty improvement}. We see that the peak lies around an improvement of roughly 3-10x, although for many planets the required improvement is much smaller. For planets where the radial velocity amplitude or transit depth are the largest contributing factor, this implies that at a reduction in their uncertainties by 3-10x, tidal effects on density will begin to become relevant and the planet can no longer safely assume to be spherical. This is shown by the grey curve in figure \ref{fig:error breakdown}, where we reduce the density error by a factor of three and show the relative value of tidal effects, which is comparable to $\sim50\%$ of the measurement error on density for $>35\%$ of planets.

We note that there is a small sample of planets for which current measurement errors are in fact less than the calculated deviation on their densities due to tidal effects (the highlighted orange part of Figure \ref{fig:density uncertainty improvement}). We report these planets in Table \ref{tab:planets with small sigmas}, sorted by the multiplicative factor by which tidal deviations outweigh measurement uncertainties. In the worst case, we find that for WASP-19 b this is almost a factor of four, implying that the reported precision on its density is significantly underestimated. Again we note that grey curve of figure \ref{fig:error breakdown} shows that after reducing the total error on density by a factor of three, we find $\sim20\%$ of our sample or almost 40 planets for which tidal variations on density would become larger than measurement errors.

\begin{table}[]
\centering
\begin{tabular}{lcccc}
\hline
\multicolumn{1}{l}{Planet Name} & \multicolumn{1}{c}{Period (days)} & \multicolumn{1}{c}{$\Delta_{\rho} / \sigma_{\rho}$}& \multicolumn{1}{c}{Reference} \\ \hline

WASP-19 b&0.8&3.8&\cite{hebb:2009}\\
HAT-P-7 b&2.2&3.6&\cite{pal:2008}\\
WASP-12 b&1.1&3.0&\cite{hebb:2009}\\
WASP-121 b&1.3&1.9&\cite{bourrier:2020}\\
WASP-4 b&1.3&1.4&\cite{bouma:2019}\\
WASP-103 b&0.9&1.4&\cite{gillon:2014b}\\
Kepler-10 b&0.8&1.4&\cite{esteves:2015}\\
K2-141 b&0.3&1.1&\cite{malavolta:2018}\\
\hline
\end{tabular}
\caption{List of planets with density uncertainties less than the potential deviation due to tidal effects.}
\label{tab:planets with small sigmas}
\end{table}


\section{Conclusions}

Using a gravitational potential framework to determine the shape of a planet under the influence of tidal distortions, we have expanded on the work of \cite{burton:2014} and calculated the amount by which such effects may bias the estimated density of an exoplanet with orbital periods of  less than three days. In comparison to an assumption of being perfectly spherical, tidal effects serve to increase the perceived volume of an exoplanet (and thus decrease its density) by an amount of up to 15$\%$ for the shortest period planets, which agrees with the values reported in \cite{burton:2014} and \cite{leconte:2011}, which reported their results as a change in effective planet radius.Similarly, \cite{akinsanmi:2018} considered a framework of a planet with ellipsoidal variations how that would affect their lightcurves and identified many of the same planets as we do in table \ref{tab:planets with small sigmas} as being those which would exhibit the strongest signal of shape deformation.

We quantify this change more precisely in terms of the semi-major axis, planet to star mass radio, and planet radius for which we are able to derive a robust relationship (eq. \ref{eq:functional form}). This allows for a rapid estimate of the magnitude of such variations, and whether or not an analysis of a planets density (and thus its internal composition) will be significantly biased by assuming the planet to be perfectly spherical. In \cite{correia:2014} a similar expression was derived through an alternate analytic consideration, including the effect of inclination as well as the fluid Love number of the planet. We find the inclination effect to alter the density perturbation by at most $2\%$ for the planets in our sample, although in most cases the effect is much less than $1\%$. The additional consideration of the fluid response of the planet implies potential variations of $\sim20\%$ between our results (i.e. a 15\% density perturbation could change by a factor of 0.8-1.2), however this is strongly affected by uncertainty in the Love number. A more detailed analysis of the fluid response of planets in \cite{wahl:2021} identified WASP-12 b, WASP-103 b, and WASP-121b as those with the potential for the greatest variation in tidal response, which we also found to be among planets with the highest deviation in derived densities.

For planets with orbital periods beyond 2.5 days we measured variations of no more than $2\%$, well below current measurement errors ($\sigma_\rho/\rho > 2.7\%$ for $p>2.5$ days). We find also that for most planets, even those with shorter orbital periods, measured uncertainties are currently too large to be affected by such deviations, however we identify a sample of planets whose uncertainties may be as much as four times smaller than the potential change caused by shape distortions. One such planet, WASP-103 b, was recently found to show tentative tidal deformations using multiple transit observations \citep{barros:2022}, where it is reported that the volumetric radius of a fit derived using an ellipsoidal planet model is 5-6$\%$ larger than the radius derived from a spherical planet model. This further strengthens the notion that for such planets susceptible to tidal deformation, any attempts to characterise their interior composition based on their density derived using spherical planet models are likely to be under-estimating their errors, and that there is a wall of accuracy which is limited by a lack of knowledge of their true shape. For other short period planets however we calculate that an overall improvement by a factor of three in density error would cause $\sim25$ planets to have density errors comparable to tidal distortions, and for $\sim50$ planets tidal distortions would compare to at least $50\%$ of the measured density uncertainty.

With this in mind, we find that radius values in a range of up to a $5\%$ deviation could in fact correspond to planets with the same density. This implies that composition curves are not just one-to-one functions but rather correspond to a family of mass-radius relationships, where there is a degeneracy induced by a lack of knowledge of the shape of a planet. This highlights a fundamental limit in the precision of characterising the composition of an exoplanet when disregarding tidal variations, which will become more severe as measurement errors decrease.

Finally, we break down the uncertainty on a planets density further as a contribution of five underlying factors, three of which are directly observable and two which are model-derived. This breakdown highlights the fact that it is the directly observable quantities (specifically RV amplitude and transit depth) which are in most cases responsible for the bulk of the error in a planets density (in some cases contributing almost the entirety of the error budget). We also find that the median contribution of the largest piece of the uncertainty budget is $59\%$, implying that for most planets there is a single key parameter contributing the bulk of the uncertainty. Thus upcoming extreme precision RV measurements as well as high SNR transit observations such as those from JWST and PLATO imply that biases due to tidally-induced shape deformations will become a significant and unavoidable bottleneck when attempting to measure the density of planet to a high level of accuracy as the error in these key contributing factors is reduced.

\section{Acknowledgements}
DB
acknowledges support from an FRQNT Doctoral Research Scholarship.

\bibliography{ms}
\end{document}